\documentclass[usegraphicx,usenatbib]{mn2e}
\usepackage{pdf14,times}
\usepackage{fixltx2e}
\usepackage{mathptmx}
\usepackage{multirow}
\usepackage{lscape}

\begin{document}

\title[Properties of AGN coronae]{Properties of AGN coronae in the \textit{NuSTAR} era}
\author[A.C. Fabian et al] {\parbox[]{6.5in}{{
      A.C.~Fabian$^1\thanks{E-mail: acf@ast.cam.ac.uk}$,
      A.~Lohfink$^1$, E.~Kara$^1$, M.L.~Parker$^1$, R. Vasudevan$^1$
      and C.S.~Reynolds$^{1,2}$
    }\\
    \footnotesize
    $^1$ Institute of Astronomy, Madingley Road, Cambridge CB3 0HA\\
    $^2$ Department of Astronomy, University of Maryland, College
    Park, MD 20742-2421, USA\\
  }}
\maketitle

  
\begin{abstract}
  The focussing optics of \textit{NuSTAR} have enabled high
  signal-to-noise spectra to be obtained from many X-ray bright Active
  Galactic Nuclei (AGN) and Galactic Black Hole Binaries
  (BHB). Spectral modelling then allows robust characterization of the
  spectral index and upper energy cutoff of the coronal power-law
  continuum, after accounting for reflection and absorption
  effects. Spectral-timing studies, such as reverberation and broad
  iron line fitting, of these sources yield coronal sizes, often
  showing them to be small and in the range of 3 to 10 gravitational
  radii in size. Our results indicate that coronae are hot and
  radiatively compact, lying close to the boundary of the region in
  the compactness -- temperature ($\Theta-\ell$) diagram which is
  forbidden due to runaway pair production. The coincidence suggests
  that pair production and annihilation are essential ingredients in
  the coronae of AGN and BHB and that they control the shape of the
  observed spectra.
\end{abstract}

\begin{keywords}
black hole physics: accretion discs, X-rays: binaries, galaxies 
\end{keywords}


\section{Introduction}
The variable hard X-ray emission from Active Galactic Nuclei is
generally considered to originate in a compact region known as the
corona which lies above the accretion disc \citep[see e.g.,][]{Vaiana1978,Haardt1993,Merloni2003}. The inner
accretion disc produces copious UV emission in a quasi-blackbody
spectral shape which is Compton upscattered by hot coronal electrons
energised by magnetic fields from the disc. Rapid variability of the
2--10~keV X-ray emission seen from many AGN indicates that the corona
is physically small. Recent X-ray spectral-timing and reverberation
analyses of AGN spectra strongly support this conclusion and in many
cases require the corona to lie just $3-10 r_{\rm g}$ above the
central black hole \citep{Fabian2009,DeMarco2011,Kara2013,Uttley2014,Cackett2014,Emmanoulopoulos2014}. 
Variability analyses of microlensed components of several lensed quasars also indicate small hard X-ray
emission regions, with half-light radii less than $6r_{\rm
  g}$. Together, the evidence points to the corona being compact
\citep{Fabian2012,Reis2013}. Further evidence of a small
physical size of the corona emerges from observations of the
emissivity profile of the broad iron line \citep{Wilkins2011} and
from varying obscuration of the corona by clouds \citep{Risaliti2011,Sanfrutos2013}.

Sources which are physically small and highly luminous can also be
compact in a radiative sense, meaning that interactions involving
significant energy exchange between photons and particles are
commonplace in the source. The relevant parameter here is then the
ratio of source luminosity to size \citep[$L/R$:][] {Cavaliere1980},
usually given in terms of the dimensionless compactness parameter
\citep[][ hereafter GFR]{Guilbert1983}:
\begin{equation}
\ell = \frac{L}{R} \frac{\sigma_{\rm T}}{m_{\rm e} c^3}
\end{equation}
$L$ is the luminosity, $R$ the radius of the source (assumed
spherical), $\sigma_{\rm T}$ the Thomson cross section and $m_{\rm e}$
the mass of the electron. When $\ell\sim 1$ a particle loses a
significant fraction of its energy on crossing the source region (see
section 2). 

An idea of the magnitude of $\ell$ can be
seen by replacing $R$ by $R_{\rm g}$, the gravitational radius
($GM/c^2$), and comparing the luminosity $L$ with the Eddington limit $L_{\rm E}$,
\begin{equation}
\ell = 4 \pi \frac{m_{\rm p}}{m_{\rm e}}\frac{R_{\rm g}}{R}\frac{L}{L_{\rm
      E}}.
\end{equation}
Thus sources operating with coronal emission exceeding 1 per cent of
the Eddington limit with coronae which are less than  $\sim 20 r_{\rm g}$ in
size are compact in the sense that $\ell>10$.

If the photons are energetic enough, photon-photon collisions can lead
to pair production which can play a major role in determining the
outgoing spectrum and overall composition of the corona
\citep{Svensson1982,Svensson1984,Guilbert1983,Zdziarski1985}. Consider
a small region of size $R$ containing soft photons and unit Thomson
depth of electrons (i.e. $\tau_{\rm T}=n_{\rm e}\sigma_{\rm T} R = 1$)
into which an increasing amount of power is fed into the electrons. As
the electron temperature $\Theta=kT_{\rm e}/m_{\rm e} c^2$ rises,
Compton scattering of the soft photons produces a power-law radiation
spectrum extending to a Wien tail at energies around
$2\Theta$. Photon-photon collisions create electron-positron pairs
when the product of the photon energies, in units of $m_{\rm e} c^2$,
exceeds 2. Thus pairs appear when the tail extends above
$\sim 2 m_{\rm e} c^2$ (i.e. 1 MeV).  The pair density is proportional
to the luminosity and temperature and inversely proportional to the
source size. The energy associated with increased luminosity goes into
increased numbers of pairs rather than temperature.  Pair production
can become a runaway process, outstripping annihilation, soaking up
energy and limiting any rise in temperature\footnote{The existence of
  a temperature limit due to pair production by particle
  collisions in a low density plasma was first shown by \citet{Bisnovatyi1971}.}. 
When the source is radiatively compact this
happens at $\Theta<0.2$ ($kT_{\rm e}<100\,{\rm keV}$).

Photon-photon collisions involving hard X-rays can thus produce
electron-positron pairs which act as an $\ell$-dependent thermostat
\citep{Svensson1984,Zdziarski1985,Pietrini1995,Stern1995,Coppi1999,Dove1997}. This limits the source temperature
$\Theta = kT_{\rm e}/m_{\rm e} c^2$ as a function of $\ell$. The only
other major factor here is the source geometry and its location with
respect to the soft photon source.

It has been suspected for over 30 years that luminous AGN and black
hole binary (BHB) sources such as Cygnus X-1 are radiatively compact.
\citet{Done1989} tabulate compactness parameters of several dozen
AGN based on their observed X-ray luminosity and an estimate/limit of source
size, obtained  from rapid
variability. They found many sources with $1<\ell<100$.  However, robust
estimates of the coronal temperature and in particular the size of the
corona have however been lacking. Now, however, with spectral
timing results combined with temperature estimates becoming available 
from \textit{NuSTAR}, robust values of both $\ell$ and $\Theta$
can be obtained, after carefully accounting for disc reflection in the 
spectrum. \textit{NuSTAR} is the first telescope for cosmic X-ray sources with
focussing optics operating in hard X-rays up to 78~keV. This means
that it yields high signal-to-noise spectra over the band where the 
corona emits most of its power.

In this paper we gather results from \textit{NuSTAR} and from the
coded mask instruments Integral and \textit{Swift}-BAT on AGN and BHB
to map out their locations on the $\ell - \Theta$ diagram. High power
levels fed into a compact corona should produce a spectrum
(e.g. photon index and high energy turnover) constrained by source
size, luminosity and geometry.  Observations of source size and
luminosity can in principle therefore enable us to establish the
geometry of the corona and reveal the heating and thermalization
mechanisms operating.

Note that we only consider coronae where emission is
generated, not the large scattering corona in the low-mass X-ray
binaries known as Accretion Disc Corona sources (e.g. \citep{Church2004}.

\section{The $\Theta - \ell$ plane}

We begin by gathering together various theoretical constraints from
GFR, \citet{Svensson1984}, and \citet{Fabian1994} within the temperature --
compactness, $\Theta - \ell$, plane.
The estimates are initially deduced for  non-relativistic temperatures,
but the coupling times are extended into the transrelativistic regime  using the
work of \citet[][ hereafter GHF]{Ghisellini1993}. 

Consider a spherical source of size $R$ and  scattering optical depth $\tau$ in
which luminosity $L$ is generated. The energy density in photons is then
\begin{equation}
 \varepsilon= \frac{L}{4 \pi R^2 c} (1+\tau)
\end{equation} 
and thus the Compton cooling time of an electron is
\begin{equation} t_C = \frac{3 \pi R}{2 c \ell (1+\tau)}. \end{equation} 
Expressed in terms of the light crossing time of the source,   $t_{\rm
  cross}$, 
\begin{equation} \frac{t_{\rm C}}{t_{\rm cross}} = \frac{3 \pi }{2 \ell (1+\tau)} \end{equation} 
If $\ell>2$
then we have ${t_{\rm C}}<t_{\rm cross}.$ This justifies the statement made in the Introduction that, when
$\ell$ exceeds unity, an electron loses much of its energy on crossing the source. 

The dominant radiation process will be the one with the shortest cooling
time. The bremsstrahlung cooling time of an electron in a gas of
density $n$ is
\begin{equation}{t_{\rm B}}=\frac{\Theta^{1\over 2}}{n\alpha_{\rm
      f}\sigma_{\rm T}c}=\frac{\Theta^{1\over 2}R}{\tau \alpha_{\rm f}
    c},
\end{equation}
where $\alpha_{\rm f}$ is the fine-structure constant, 
which we compare with $t_{\rm C}$,
to give 
 \begin{equation} \ell= \frac{3 \pi \alpha_{\rm f}}{2 \Theta^{1\over 2}} \frac{\tau}{1+\tau} 
\end{equation}
and an upper limit to bremsstrahlung dominance when 
 \begin{equation} \ell\approx 3 \alpha_{\rm f} \Theta^{-{1\over 2}}. \end{equation}
This demonstrates that Comptonization dominates at high compactness.

Two-body collisions are the simplest heating and thermalization
mechanism, with electron-proton coupling occurring faster than
electron cooling when
\begin{equation} \ell<0.04 \Theta^{-3/2}; \end{equation}
 the relevant relation for  electron-electron coupling is 
 \begin{equation} \ell<80 \Theta^{-3/2} \end{equation}
 \citep{Fabian1994}.  Both of these relations are non-relativistic and 
 apply only at low electron temperatures ($\Theta<<1$): in Fig.~1, we plot the
 electron-proton and electron-electron coupling lines from GHF, which
 have been calculated through the transrelativistic regime.

Pair production at high $\ell$ values is dominated by photon-photon
collisions. An estimation of the pair density requires a detailed
calculation covering pair production and annihilation and also the thermal
balance. Above a certain level, which we call the pair line,
pair production runs away, as described in the Introduction.    
\citet{Svensson1984} estimated that, for an isolated cloud, the pair
balance line occurs where 
\begin{equation}\ell\sim 10\Theta^{5/2}\exp(1/\Theta).\end{equation}

The position of the pair line depends on the geometry of the region and
the origin and level of the dominant soft photon field. \citet{Stern1995}
computed the pair balance curve for a slab corona above a reflecting
disc and obtained the line included in Fig.~1. They also computed the
constraints for a
hemispherical corona on a slab and a sphere at a height equal to half
the radius of the sphere \citep[see][]{Svensson1996}.

\begin{figure}
  \centering
  \includegraphics[width=0.99\columnwidth,angle=0]{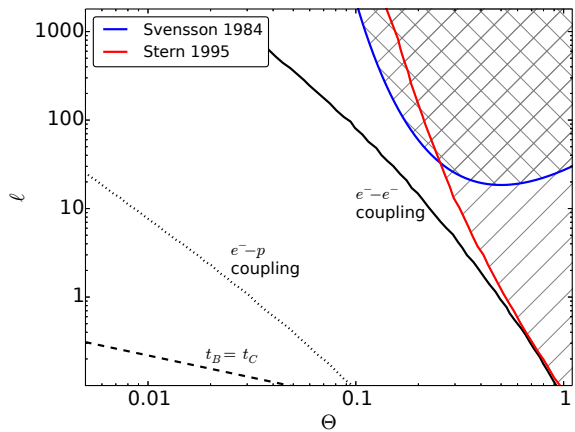}
  \caption{Summary of our theoretical understanding of the
    $\Theta-\ell$ plane as decribed in detail Section 2, included are
    the boundaries for electron-electron coupling, electron-proton
    coupling, the dominance of Compton cooling and pair lines for
    different assumptions.}
\end{figure}

\begin{figure}
  \centering
  \includegraphics[width=0.99\columnwidth,angle=0]{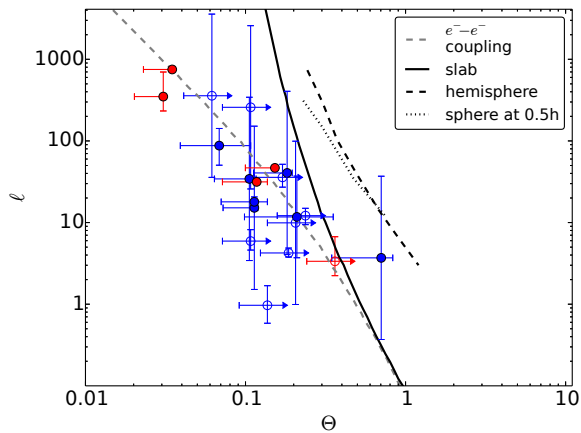}
  \caption{$\Theta-\ell$ distribution for \textit{NuSTAR} observed AGN (blue
    points) and BHB (red points). The e--e coupling line from GHF is
    included. Pair lines from Stern et al (1995) are shown. The slab
    line has been extrapolated slightly to higher $\ell$. }
\end{figure}

\section{The location of observed sources on the $\Theta-\ell$ plane}

In the last section we have seen what physical processes govern the
placement of observations in the $\Theta-\ell$ plane. This section
concerns the collection and analysis of the data used to populate the
observational $\Theta-\ell$ plane presented in this paper.

\subsection{From observations to the $\Theta-\ell$ plane}

To construct our observational $\Theta-\ell$ plane, the coronal
temperature ($kT_e$) is estimated throughout this work from the high
energy cut-off ($M(E)\sim E^{-\Gamma}\,\exp(-E/E_{\rm cut})$).  To
convert from the cut-off energy to the coronal temperature a factor of
2 is used ($kT_e=E_{\rm cut}/2$). As discussed in \citet{Petrucci2001}
a conversion factor of 2 is appropriate for optical depths smaller
than one; if the optical depth is much larger than one, the factor is
closer to 3. We account for the uncertainty of the conversion in the
errors on $\Theta$ in addition to the statistical uncertainty. For
determining luminosities, the flux of the power-law component
in the 0.1--200 keV band is estimated and converted using a value of 68\,km/s/Mpc
is used for the Hubble constant.

\subsection{Sample Selection and Data Collection}

The data we will consider is divided into two categories; the first
is where the information on the high energy cut-off is from a
\textit{NuSTAR} observation (CAT 1) and the second is where the
information on the high-energy cut-off is from the BAT instrument
onboard \textit{Swift} (CAT 2). Finally, we include results obtained
from microlensed quasars observed with \textit{Chandra} in the second
category.

In our CAT 1 sample we include all those sources which have work which
is published or in preparation that provide a constraint on the
high-energy cut-off from \textit{NuSTAR}, even if it is only a lower
limit. The so-selected CAT 1 targets are shown, together with the
measurements of our observables, in Tables \ref{data_nustar}, and
\ref{data_binaries} for AGN and BHB, respectively. (Most of the BHB 
are in the hard state.) As \textit{NuSTAR}
provides spectra with the best signal-to-noise in the hard X-ray band
and modelling usually includes detailed fitting of any reflection
components, we expect \textit{NuSTAR} spectra to provide the most
robust results.

The values for the high energy cut-offs of the CAT 1 sample stem from
the individual works quoted in the right-most column of the tables. If
more than one epoch of observations exists we select the one with the
best constraint on the cut-off. The estimates on the coronal size
originate from modeling of the reflection spectrum and X-ray
reverberation measurements (unless the source is lensed). If no
measurement exists we assume a value of $10\,R_{\rm g}$ which is a
conservative assumption given the measurements. For the black hole
masses required to derive the actual size of the coronal region, we
use optical broad line reverberation measurements where they are
available. If the mass estimate in the literature does not have an
error estimate we assume a conservative uncertainty of a factor 10 for
AGN. For Galactic black hole binaries we assume a black hole mass of
$10\pm5\,M_\odot$ if no mass estimate exists. For sources observed
with \textit{NuSTAR}, we estimate the coronal luminosities from the
coronal fluxes of the best fit spectral models presented in the
individual works. To do this we use a cut-off power law with their
best fit $\Gamma$, $E_{\rm cut}$, and power law normalization, if
given. If no power law normalization is given we determine it from the
flux measurement of the most power law dominated band, given in the
paper. The coronal flux is then determined in the energy range
0.1-200\,keV, as below this band the spectrum could deviate
significantly from a power law and most sources already roll-over
below 200\,keV. In case of the X-ray binaries we then use the distance
to get the luminosity and for AGNs we use the redshifts, which are
drawn from the NASA/IPAC Extragalactic Database (NED).

Our CAT 2 sample is assembled from three works; an analysis of
XMM/INTEGRAL/BAT Type 1 AGN observations by \citet{Malizia2014a}, an
analysis of XMM/BAT AGN observations in the Northern Galactic Cap by
\citet{Vasudevan2013} and a microlensing sample presented in
\citet{Chen2012}.  In the first case the BAT spectra are those of the
70 month catalogue, and in the second case those of the 58 month
catalogue. We select from the first two samples, subsamples of sources
with mass estimates and constraints on the high energy cut-off. For
the microlensing sample we use sources where an estimate of the
coronal size exists. The selected sources and measurements are listed
in Tables \ref{data_integral}, \ref{data_ranjan}, and
\ref{data_sample_lenses}. 

The assumptions made for the coronal temperatures, coronal size and
black hole mass are the same as for the CAT 1 sample. The coronal
fluxes for the \citet{Vasudevan2013} sources are also determined in a
similar fashion to what was described above. For the
\citet{Malizia2014a} sample the same method cannot be applied as no
flux values are given in the paper, instead we use the best fit BAT
cross-calibration constant. We downloaded the 70 month BAT spectra of
the sources and set up a spectral model consisting of a cut-off power
law multiplied by the cross calibration constant. The photon index,
high energy cut-off and cross calibration constant are kept fixed at
their best fit values, while we fit for the normalization. Once the
best fit normalization is found we delete the constant and estimate
the coronal flux in the 0.1-200\,keV just as before. Again, the
redshifts from NED are used to obtain the coronal luminosity. The
microlensed sample is treated differently as the measurement of the
physical size of the X-ray emitting region is already
non-dimensionless and can be used to calculate $\ell$ directly, the
black hole mass is therefore not required. We assume further the
highest energy bin in the \textit{Chandra} spectra shown in
\citet{Chen2012} is a lower limit on the high-energy cut-off. This is
them corrected for redshift. 
Finally, for the coronal luminosities of the lensed quasars we use the
0.2-50\,keV luminosities reported in the \citet{Chen2012} paper. These
objects serve to show that distant quasars operate at values of $\ell$
within the range 10--100 or greater; NuSTAR observations will
be required to measure their expected cutoffs and thus $\Theta$ values.

\section{Results}

The results for the \textit{NuSTAR} sample (CAT1) are plotted in
Fig.~2. Although some of the results are lower limits, about one half
are measurements of clear turnovers seen in the \textit{NuSTAR} band. A few examples
of AGN with clear turnovers are MCG-5-23-16 and SWIFT~J2127.4+5654,
while examples of  BHB are GRS~1915+105, GRS~1739-278 and Cyg X-1. The
largest cutoff energy is inferred in the AGN NGC5506, which is at low
$\ell$.

\begin{figure}
  \centering
  \includegraphics[width=0.99\columnwidth,angle=0]{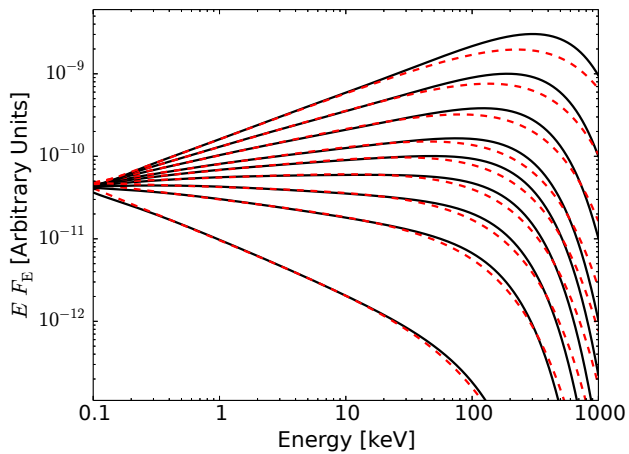}
  \caption{ COMPPS (black) versus exponentially cutoff power-law
    (red). See text for details. }
\end{figure}

Several factors can affect the measurement of coronal temperature. The
first is observational and due to the common useage of an exponential
cutoff as a model for the upper spectral turnover produced by
Comptonization. As shown by \citet{Zdziarski2003}, this produces a
slower break than Comptonization, which retains a straight power-law
shape to higher energies before more abruptly turning down. We have
simulated this by generating Comptonization spectra using COMPPS
\citep{Poutanen1996} in a spherical geometry for a seed photon
tenperature of 10\,eV, a Thomson depth $\tau=1$ and a set of
temperatures (40\,keV, 60\,keV, 70\,keV, 80\,keV, 90\,keV, 100\,keV,
120\,keV, 150\,keV, 200\,keV) (Fig.~3). These have been fitted with a
powerlaw in the 0.5--10~keV band and the power law was later modified at
higher energies using an exponential turnover assumed to represent the
effect of a turnover at $2\,kT_{\rm e}$.  The plot demonstrates that when the
temperature is significantly above the \textit{NuSTAR} band
(i.e. $\Theta>0.2$) and no significant deviation is yet seen in the
\textit{NuSTAR} spectrum, then it is possible that the cutoff energy
and thus temperature are overestimated.  These affects have been
carefully considered by \citet{Matt2015} in their study of the
\textit{NuSTAR} data on NGC5506. The exponential cutoff energy is
measured at $720^{+130}_{-190}\,{\rm keV}$ whereas the COMPPS model
\citep{Poutanen1996} gives a temperature of 270~keV. \citet{Matt2015}
also remind us of the comment by \citet{Gilli2007} that the mean value
of $E_{\rm cut}$ for AGN must lie below 300~keV, in order not to
saturate the X-ray Background at 100~keV. Objects such as NGC5506
must be exceptional rather than the norm.

\begin{figure}
  \centering
  \includegraphics[width=0.99\columnwidth,angle=0]{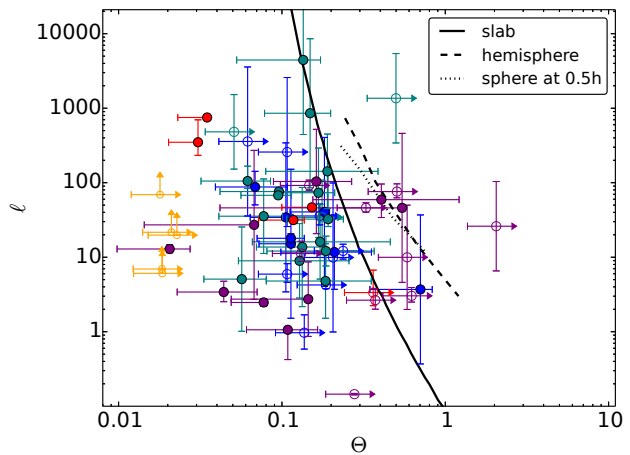}
  \caption{All measurements including Integral (teal), \textit{Swift}-BAT
    (magenta) and gravitational lensed objects (yellow).}
\end{figure}

The results including the CAT2 sample are plotted in Fig.~4. As mentioned
already these generally encompass cruder spectral modelling than used
for the \textit{NuSTAR} data, and the total exposures were each accumulated over
several years. The sources generally concentrate around similar values
to those measured with \textit{NuSTAR}, although there are a few outliers at
high temperature. The most extreme one with $\Theta\sim2$ is NGC4388,
a Seyfert 2 galaxy in the Virgo cluster. The measurement is with the
Swift-BAT, the upper energy of which is 200~keV so could be suffering
from the effect shown in Fig.~ 4. Objects where there are two pairs of
$\Theta-\ell$ values are plotted in Fig.~5. 

\begin{figure}
  \centering
  \includegraphics[width=0.99\columnwidth,angle=0]{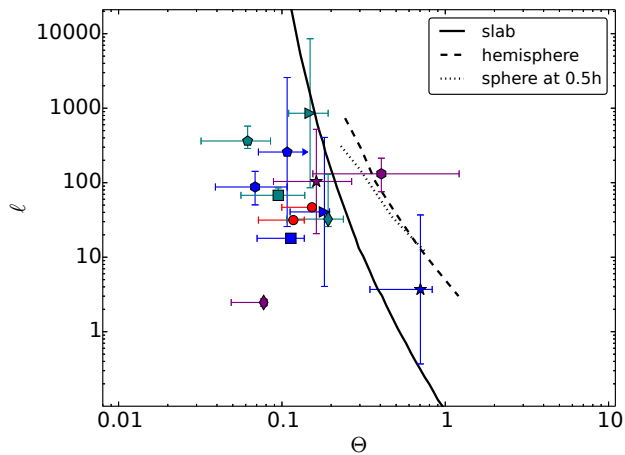}
  \caption{Distribution of objects which have more measurements of
    $\ell$ and $\Theta$ from different instruments. The two red points
  are the two states of Cyg X-1.}  
\end{figure}

The distribution of high energy cutoff values from the various samples
is shown in Fig.~6. To make use of all measurements including lower
limits,  we calculate the histograms and errors using Monte-Carlo
simulations. For each realization of the histogram we draw for each
cutoff constraint a cutoff value within its error range (for the lower
limits, 1000~keV is assumed to be the upper limit.) The histograms
show clear peaks in the range of 100--150~keV.

\begin{figure}
  \centering
  \includegraphics[width=0.99\columnwidth,angle=0]{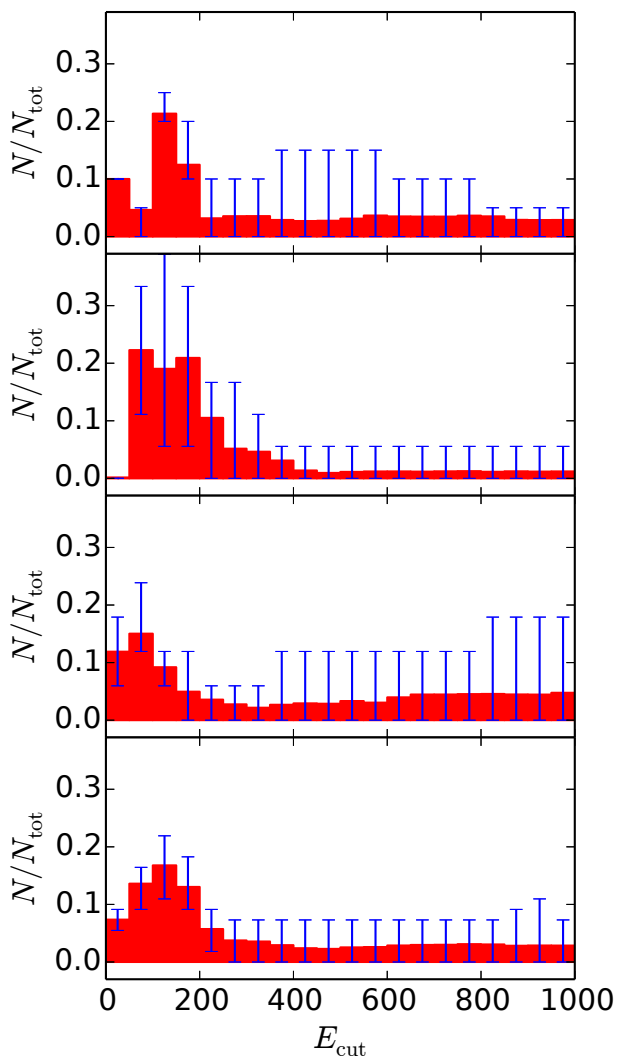}
  \caption{Histograms of $\Theta$ estimates: from top to bottom,
    NuSTAR sample (see text), Integral Malizia et al (2014), Swift-BAT
    (Vasudeven et al (2013), all.  }
\end{figure}

\section{Discussion}

We find (Fig.~2) that the X-ray coronae measured by \textit{NuSTAR}
have $\ell$ ranging from 1 to 1000 with most lying between 10 and 100,
and $\Theta$ ranging from 0.03 to 0.8, with most between 0.07 and
0.3. This places many sources against the pair runaway line for slab
geometry. Given the remaining uncertainties in spectral modelling and
size estimates it is plausible that many of the sources track close to
that line. The clustering of sources in the $\Theta-\ell$ plane argues
for a general physical reason, such as the pair line, attracting or
constraining them to that region.

Many of the coronae lie in the region where General Relativistic
effects such as gravitational redshift and light-bending are
important. To see one change produced by the first effect, the
temperatures have been corrected for gravitational redshift $z$ in
Fig.~7. The points now lie closer to the pair line for a slab but
several still lie at a  significantly lower value of $\Theta$.  GR will
also affect our estimates of $\ell$ through the effects on $L$
($(1+z)^{(3+\alpha)}$ from Liouville's theorem) and $R$ (which depends
on the method by which this has been inferred). Light bending, in
which the strong gravity bends radiation from the corona down towards
the disc, requires another correction which is inclination
dependent. The net effect is to boost the intrinsic values of $\ell$
by factors of around 2--10 above the observed estimates. As there is
more uncertainty in these corrections we merely note that the
estimates of $\ell$ should be seen as lower limits.

The coronal emission is likely anisotropic when the corona is close to
the centre of the disc with factors such as spin and inclination 
affecting appearance. \citet{Wilkins2011} have used the
emissivity profile of the iron line in several AGN to deduce the
coronal illumination of the disc and thus indirectly constrain the
size of the corona. In several cases the bulk of the coronal emission
is tightly limited to a compact region above the centre of the disc,
with radius $r$ less than the height of the corona $h$, i.e. $r<h$,
but with some emission originating from larger radii at the same
height. 

The geometrical considerations in the Stern et al (1995) models mainly
affect the ratio of soft photon to electron heating powers in the
corona.  The effects of light bending on the incidence of soft
blackbody disc emission on the corona and the twisting of magnetic
fields from the disc can modify the relevance of those geometries to
the strong gravity regime above the centre of the accretion disc. As
seen from the corona, the disc will appear to cover most of the Sky,
possibly intensifying the effects of Compton cooling and moving the
pair line to lower $\Theta$.

\begin{figure}
  \centering
  \includegraphics[width=0.99\columnwidth,angle=0]{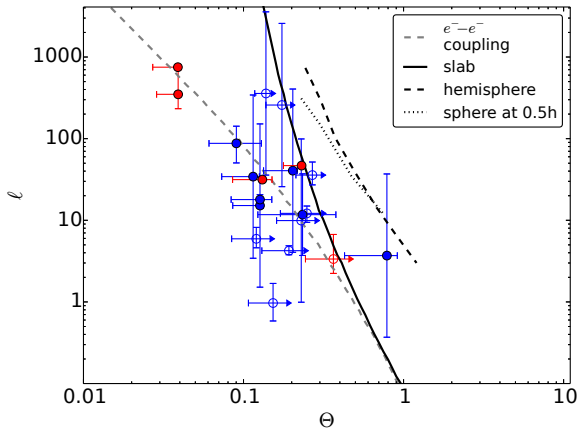}
  \caption{NuSTAR $\Theta-\ell$ distribution corrected for gravitational redshift.   }
\end{figure}

Another possible reason for the offset in $\Theta$ values from the
pair line is our assumption that the corona is homogeneous and single
temperature. Instead, the corona may be very dynamic with heating
localised and highly intermittent in space and time. This could
produce a range of temperatures in the corona, all limited by the pair
threshold, but with the mean at a lower value due to Compton
cooling. Since pairs are produced by the high energy tail of the
particle and photon energy distributions, then only a small disperson
in temperatures can have a large effect. This would imply that deeper
observations\footnote{We note that the BHB GRS~1915+105
  \citep{Miller2013}, GRS~1739-278 \citep{Miller2015} and Cyg X-1
  (Parker et al 2015) already show signs of possible hard tails in
  their spectral residuals.}  should reveal a high energy tail to the
spectra above that expected from single temperature Comptonization.

A variant of this possibility occurs if the protons in the corona are
much hotter and the source of some of the electron heating, then
proton-electron coupling may create a high energy tail which produces
pairs at a lower ``temperature'' than expected for a strict Maxwellian
distribution. More generally, such effects will apply to sources above
the electron-electron coupling line. Indeed, the parts of the corona of
where heating is most intense may exceed this constraint. The electron
distribution may not then be Maxwellian. This need not greatly affect
the Comptonized spectrum, which will resemble that from a thermal
electron population with the same mean value of $(\gamma^2-1),$ where
$\gamma$ is the electron Lorentz factor \citep[GHF; see also][]{Nayakshin1998}.

Many sources are (marginally) above the electron--electron coupling
line and all are above the electron--proton line. This emphasizes the
long standing problems of the heating and thermalization process
\citep{Guilbert1982, Ghisellini1993,Svensson1999,Merloni2001}.  All
sources have a cooling time less than the light crossing time so the
energy must be present there in some other form. Since protons cannot
supply the energy to the electrons fast enough, due to the long
coupling time (Fig.~1) we presume that it must be in terms of magnetic
field. As indicated many times before \citep[e.g.][]{Merloni2001}, the
corona must be magnetically dominated. An underlying strong magnetic
field means that the synchrotron boiler dominates the energy exchange
between low-energy electrons and the photon field
\citep{Ghisellini1988,Ghisellini1998,Belmont2008,Veledina2011}. Taking 
these issues further is beyond the scope of this paper.

We note that all the sources that we have examined are above the
Eddington limit for pairs. We have assumed in all cases that the 
source is static. Magnetic fields are presumably responsible for
holding the bulk of the source together, although that does not
exclude a pair wind escaping from part of the source
\citep[e.g.][]{Beloborodov1999} or being accelerated into a jet 
\citep{Henri1991,Moscibrodska2011}. The pair limit may be avoided 
if the emission region is part of a relativistic outflow, such as a 
Gamma-ray burst (e.g. Piran 2004). Note that no distinct annihilation
line should be observable from a thermal pair plasma
\citep{Zdziarski1984}.

The modelling by \citet{Stern1995} and others indicates that the
Thomson depth of the scattering region should be less than unity. Most
of the coronal particles can be pairs. Pair balance together with the
geometry dictate the expected epectral shape of the coronal
continuum. The observed photon index $\Gamma$ is plotted against
$\ell$ in Fig.~8.  There is some agreement with the \citet{Stern1995}
model predictions in the $\Gamma -\ell$ plane. We also include a line
from the work of \citet{Shemmer2008} who find an observational
correlation (with much scatter) between Eddington fraction, $L/L_{\rm
  Edd}$, and  photon index $\Gamma$. For a given coronal size $R$,
this relates to $\ell$ through equation (2). The position of this line
in the y-axis thus depends on $R$. 

We stress that the pair limit to the temperature of luminous, static, compact
regions cannot be avoided. It explains the rough uniformity and trend
in cutoff energy (higher values of $E_{\rm cut}$ occur at lower values
of $\ell$ in Fig.~2) that is emerging from \textit{NuSTAR}
observations of AGN and BHB. Temperatures close to the pair line at
$\Theta\sim 0.1-0.25$ are expected when such high powers are
dissipated in the physically compact regions found immediately around
accreting black holes. Further detailed computations are required to
make more precise predictions for sources in the observed
$\Theta - \ell$ plane, taking into account the effects of heating,
thermalization, inhomogeneities, geometry and light bending.  Future
hard X-ray observations with NuSTAR and ASTRO-H will map the
$\Theta-\ell$ plane in more detail and, with improved theoretical
models, lead to a deeper understanding of the central engine of AGN,
the most luminous persistent sources in the Universe.

\begin{figure}
  \centering
  \includegraphics[width=0.99\columnwidth,angle=0]{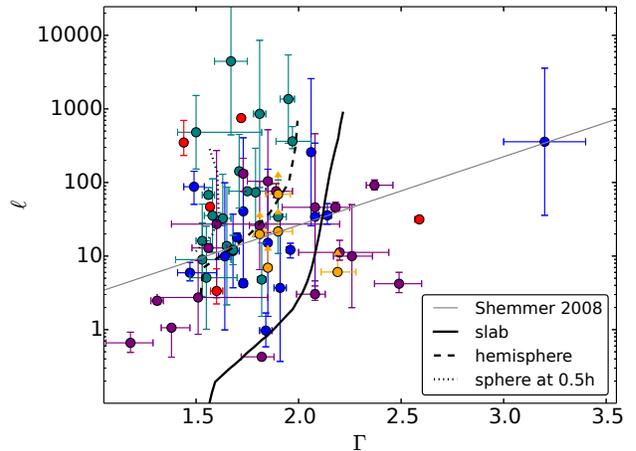}
  \caption{Distribution of $\ell$ plotted versus photon indices
    $\Gamma$. Lines from the Stern et al (1995) models (see also
    Svensson 1996) are indicated.   }
\end{figure}

\section*{Acknowledgements}

We thank Julien Malzac for a discussion and the referee for helpful comments. 
ACF and AL acknowledge support from ERC Advanced Grant FEEDBACK.  CSR
thanks the Simons Foundation Fellows Program (US) and the Sackler
Fellowship Program (Cambridge) for support, and is grateful to the
Institute of Astronomy (Cambridge) for its hospitality during a six
month visit in which this work was conducted.

\bibliographystyle{mn2e}

\section{Appendix: Data}

\begin{table*}
\caption{The targets and properties of the active galactic nuclei with cut-off constraints resulting from observations with \textit{NuSTAR}. The references to the individual works are given in the right-most column.}\label{data_nustar}
\begin{center}
\begin{tabular}{|c|c|c|c|c|c|c|c|c|c|c|}
\hline
Source & z & $\log(M)$ & $r_{\rm co}$ & $F_x$ & $E_{\rm cut}$ & $\Gamma$ & $\Theta$ & $\ell$ & Data & References \\
       &   &  [$M_\odot$] & [$r_G$] &  & [keV] & & & \\
\hline \hline NGC\,5506 & 0.006 & $8\pm1$ & 10 & 2.9 & 720$_{-190}^{+130}$ & $1.91_{-0.03}^{+0.03}$ & 0.71$_{-0.36}^{+0.13}$ & 4$_{-3}^{+33}$ & SWIFT/NU &  $1-2$ \\
NGC7213 & 0.006 & $7.98_{-0.24}^{+0.22}$ & 10 & 0.71 & $>240$ & $1.84_{-0.03}^{+0.03}$ & $>0.05$ & 1.0$_{-0.4}^{+0.7}$ & NU &  $3-4$ \\ 
MCG-6-30-15 & 0.008 & $6.7\pm1$ & 2.9 & 8.2 & $>110$ & $2.061_{-0.005}^{+0.005}$ & $>0.04$ & 258$_{-232}^{+2323}$ & XMM/NU & $5-6$ \\
NGC\,2110 & 0.008 & $8.3\pm1$ & 10 & 8.9 & $>210$ & $1.64_{-0.03}^{+0.03}$ & $>0.07$ & 10$_{-9}^{+89}$ & SWIFT/NU & $7-8$ \\
MCG 5-23-16 & 0.009 & $7.85\pm1$ & 10 & 4.2 & 116$_{-5}^{+6}$ & $1.85_{-0.01}^{+0.01}$ & 0.11$_{-0.04}^{+0.01}$ & 15$_{-14}^{+136}$ & NU & $9-11$ \\ 
SWIFT J2127.4+5654 & 0.014 & $7.18\pm1$ & 13 & 1.1 & 108$_{-10}^{+11}$ & $2.08_{-0.01}^{+0.01}$ & 0.11$_{-0.04}^{+0.01}$ & 34$_{-31}^{+308}$ & XMM/NU & $12-13$\\
IC4329A & 0.016 & $8.1\pm1$ & 10 & 4.9 & $186_{-14}^{+14}$ & $1.73_{-0.01}^{+0.01}$ & 0.18$_{-0.07}^{+0.01}$ & 41$_{-37}^{+365}$ & SU/NU & $14-15$ \\
NGC 5548 & 0.018 & $7.59_{-0.21}^{+0.24}$ & 4.5 & 1.3 & 70$_{-10}^{+40}$ & $1.49_{-0.05}^{+0.05}$ & 0.07$_{-0.03}^{+0.04}$ & 88$_{-37}^{+55}$ & XMM/NU & $5,16-17$\\
Mrk\,335 & 0.026 & $7.42_{-0.16}^{+0.12}$ & 3 & 0.10 & $>174$ & $2.14_{-0.04}^{+0.02}$ & $>0.06$ & 36$_{-9}^{+16}$ & SWIFT/NU & $18-19$ \\
Ark\,120 & 0.033 & $7.66_{-0.06}^{+0.05}$ & 4.4 & 0.55 &  $>68$ & $1.73_{-0.02}^{+0.02}$ & $>0.06$ & 4$_{-1}^{+1}$ & XMM/NU & $20-21$ \\
1H0707-495 & 0.041 & $6.31\pm1$ & 2 & 0.14 & $>63$ & $3.2_{-0.2}^{+0.2}$ & $>0.02$ & 358$_{-322}^{+3219}$ & SWIFT/NU & $22-23$\\
Fairall~9 & 0.047 & $8.41_{-0.09}^{+0.11}$ & 21 & 0.87 & $>242$ & $1.96_{-0.02}^{+0.01}$ & $>0.08$ & 12$_{-3}^{+3}$ & XMM/NU & $20,24$ \\
3C390.3 & 0.056 & $9.40_{-0.06}^{+0.05}$ & 10 & 1.6 & 116$_{-8}^{+24}$ & $1.70_{-0.01}^{+0.01}$ & 0.11$_{-0.04}^{+0.02}$ & 18$_{-2}^{+3}$ & SU/NU & $25-26$\\
Cyg\,A & 0.056 & $9.40_{-0.14}^{+0.11}$ & 10 & 1.1 & $>110$ & $1.47_{-0.06}^{+0.13}$ & $>0.04$ & 6$_{-1}^{+2}$ & NU & $27-28$ \\
3C382 & 0.058 & $9.2\pm0.5$ & 10 & 1.4 & 214$_{-63}^{+147}$ & $1.68_{-0.02}^{+0.03}$ & 0.21$_{-0.11}^{+0.14}$ & 12$_{-8}^{+25}$ & SWIFT/NU & $29-30$ \\
\hline
 \end{tabular}
\end{center}
\begin{flushleft}
$F_x$ is the 0.1-200\,keV X-ray flux in $10^{-10}\,{\rm erg}\,{\rm
  cm}^{-2}\,{\rm s}^{-1}$.\\
\textbf{References:} 1:\citet{Guainazzi2010}, 2:\citet{Matt2015}, 3:\citet{Ursini2015a}, 4:\citet{Blank2005}, 5:\citet{Emmanoulopoulos2014}, 6:\citet{Marinucci2014a}, 7:\citet{Moran2007}, 8:\citet{Marinucci2014b}, 9:\citet{Ponti2012}, 10:\citet{Zoghbi2014}, 11:\citet{Balokovic2015}, 12:\citet{Malizia2008}, 13:\citet{Marinucci2014}, 14:\citet{Bianchi2009}, 15:\citet{Brenneman2014}, 16:\citet{Pancoast2014}, 17:\citet{Ursini2015}, 18:\citet{Grier2012}, 19:\citet{Parker2014}, 20:\citet{Peterson2004}, 21:\citet{Matt2014}, 22:\citet{Bian2003}, 23:\citet{Kara2015}, 24:\citet{Lohfink2015}, 25:\citet{Grier2013}, 26:\citet{Lohfink2015a},
 27:\citet{Tadhunter2003}, 28:\citet{Reynolds2015}, 29:\citet{Winter2009}, 30:\citet{Ballantyne2014} 
\end{flushleft}
\end{table*}

\begin{table*}
\caption{The targets and properties of the black hole X-ray binaries with cut-off constraints resulting from observations with \textit{NuSTAR}. The references to the individual works are given in the right-most column.}\label{data_binaries}
\begin{center}
\begin{tabular}{|c|c|c|c|c|c|c|c|c|c|c|}
\hline
Source & $d$ & $M$ & $r_{\rm co}$ & $F_x$ & $E_{\rm cut}$ & $\Gamma$ & $\Theta$ & $\ell$ & Data & References \\
       & [kpc]  &  [$M_\odot$] & [$r_G$] &  & [keV] & & &  \\
\hline \hline GRS\,1739-278 & 8.5 & $10\pm5$ & 5 & 1.1 & $31.3_{-0.3}^{+0.3}$ & $1.44\pm 0.01$ & 0.0306$_{-0.0104}^{+0.0003}$ & 349$_{-116}^{+349}$ & NU & $1-3$ \\
GRS\,1915+105 & 11.0 & $10.1\pm0.6$ & 10 & 2.9 & $35.6_{-0.3}^{+0.3}$ & $1.72\pm 0.02$ & 0.0348$_{-0.0118}^{+0.0003}$ & 751$_{-42}^{+47}$ & NU & $4-5$ \\
GX\,339-4 & 8.0 & $10\pm5$ & 150 & 0.36 & $>370$ & $1.60\pm 0.03$ & $>0.12$ & 3$_{-1}^{+3}$ & XMM/NU & $6-7$ \\
Cyg X-1 soft & 1.86 & $14.8\pm1.0$ & 10 & 6.1 & $120_{-10}^{+20}$ & $2.59_{-0.02}^{+0.01}$ & 0.12$_{-0.05}^{+0.02}$ & 32$_{-2}^{+2}$ & SU/NU & $8-10$ \\
Cyg X-1 hard & 1.86 & $14.8\pm1.0$ & 3.3 & 3.0 & $156\pm 3$ & $1.568\pm 0.005$ & 0.153$_{-0.053}^{+0.003}$ & 47$_{-3}^{+3}$ & SU/NU & $8-9,12$ \\
\hline
 \end{tabular}
\end{center}
\begin{flushleft}
$F_x$ is the 0.1-200\,keV X-ray flux in $10^{-8}\,{\rm erg}\,{\rm
  cm}^{-2}\,{\rm s}^{-1}$.\\
\textbf{References:} 1:\citet{Dennerl1996}, 2:\citet{Greiner1996}, 3:\citet{Miller2015}, 4:\citet{Miller2013}, 5:\citet{Steeghs2013}, 6:\citet{Zdziarski2004}, 7:\citet{Fuerst2015}, 8:\citet{Orosz2011}, 9:\citet{Reid2011}, 10:\citet{Tomsick2014}, 11:\citet{Parker2015}
\end{flushleft} 
\end{table*}

\begin{table*}
\caption{The targets and properties of the active galactic nuclei analyzed by \citet{Malizia2014a} included in this work. The reference to the works providing the mass estimates are given in the right-most column.}\label{data_integral}
\begin{center}
\begin{tabular}{|c|c|c|c|c|c|c|c|c|c|c|}
\hline
Source & z & $\log(M)$ & $r_{\rm co}$ & $F_x$ & $E_{\rm cut}$ & $\Gamma$ & $\Theta$ & $\ell$ & Data & References \\
       &   &  [$M_\odot$] & [$r_G$] &  & [keV] & & & \\
\hline \hline
IGR\,J0033+6122 & 0.105 & $8.5\pm0.5$ & 10 & 3.2 & $>52$ & $1.50_{-0.09}^{+0.32}$ & $>0.02$ & 481$_{-329}^{+1041}$ & XMM/INT/BAT & 1\\  
3C\,111 & 0.049 & $9.6\pm0.8$ & 10 & 5.4 & $136_{-29}^{+47}$ & $1.65_{-0.02}^{+0.04}$ & 0.13$_{-0.06}^{+0.05}$ & 14$_{-12}^{+73}$ & XMM/INT/BAT & 2 \\
MCG+08-11-011 & 0.021 & $8.1\pm0.6$ & 10 & 5.1 & $171_{-30}^{+44}$ & $1.79\pm 0.01$ & 0.17$_{-0.08}^{+0.04}$ & 73$_{-55}^{+219}$ & XMM/INT/BAT & 2 \\
Mrk\,6 & 0.019 & $8.2\pm0.5$ & 10 & 0.94 & $131_{-48}^{+132}$ & $1.53_{-0.13}^{+0.14}$ & 0.13$_{-0.07}^{+0.13}$ & 9$_{-6}^{+19}$ & XMM/INT/BAT & 3 \\
IGR\,J07597-3842 & 0.040 & $8.3\pm0.5$ & 10 & 1.0 & $79_{-16}^{+24}$& $1.58\pm 0.04$ & 0.08$_{-0.04}^{+0.02}$ & 36$_{-24}^{+77}$ & XMM/INT/BAT & 4 \\
NGC\,3783 & 0.010 & $7.47_{-0.009}^{+0.007}$ & 10 & 5.6 & $98_{-34}^{+79}$& $1.75\pm 0.09$ & 0.10$_{-0.05}^{+0.08}$ & 76$_{-1}^{+2}$ & XMM/INT/BAT & 5 \\
NGC\,4151 & 0.003 & $7.5_{-0.6}^{+0.1}$ & 10 & 22 & $196_{-32}^{+47}$& $1.63\pm 0.04$ & 0.19$_{-0.09}^{+0.05}$ & 33$_{-7}^{+97}$ & XMM/INT/BAT & 6 \\
IGR\,J12415-5750 & 0.024 & $8.0\pm0.5$ & 10 & 0.64 & $175_{-74}^{+296}$& $1.53_{-0.03}^{+0.04}$ & 0.17$_{-0.11}^{+0.29}$ & 16$_{-11}^{+35}$ & XMM/INT/BAT & 1 \\
MCG-06-30-15 & 0.008 & $6.7_{-0.2}^{+0.1}$ & 2.9 & 2.1 & $63_{-14}^{+24}$&  $1.97_{-0.08}^{+0.09}$ & 0.06$_{-0.03}^{+0.02}$ & 105$_{-22}^{+62}$ & XMM/INT/BAT & $7-8$\\
IC4329A & 0.016 & $7\pm1$ & 10 & 7.7 & $152_{-32}^{+51}$& $1.81\pm0.03$ & 0.15$_{-0.07}^{+0.05}$ & 856$_{-770}^{+7701}$ & XMM/INT/BAT & 5 \\
IGR\,J16558-5203 & 0.054 & $7.9\pm0.5$ & 10 & 0.91 & $194_{-72}^{+202}$& 1.71f & 0.19$_{-0.11}^{+0.20}$ & 142$_{-97}^{+308}$ & XMM/INT/BAT & 4 \\
GRS\,1734-292 & 0.021 & $8.9\pm0.7$ & 10 & 2.1 & $58_{-7}^{+24}$& $1.55_{-0.08}^{+0.15}$ & 0.06$_{-0.02}^{+0.02}$ & 5$_{-4}^{+20}$ & XMM/INT/BAT & 9 \\
3C\,390.3 & 0.056 & $8.46_{-0.1}^{+0.009}$ & 10 & 1.5 & $97_{-11}^{+20}$& $1.56_{-0.03}^{+0.03}$ & 0.10$_{-0.04}^{+0.02}$ & 68$_{-1}^{+18}$ & XMM/INT/BAT & 5 \\
NGC\,6814 & 0.005 & $7.1\pm0.2$ & 10 & 1.3 & $190_{-66}^{+185}$ & $1.68\pm0.02$ & 0.19$_{-0.11}^{+0.18}$ & 12$_{-5}^{+7}$ & XMM/INT/BAT & 6\\
4C\,74.24 & 0.104 & $9.6\pm0.5$ & 10 & 0.41 & $189_{-66}^{+171}$ & $1.82\pm0.02$ & 0.19$_{-0.11}^{+0.17}$ & 5$_{-3}^{+10}$ & XMM/INT/BAT & 10 \\
S5\,2116+81 & 0.086 & $8.8\pm0.5$ & 10 & 0.68 & $>180$ & $1.90\pm0.04$ & $>0.06$ & 34$_{-23}^{+74}$ & XMM/INT/BAT & 3 \\
1H2251-179 & 0.064 & $6.9\pm1.0$& 10 & 2.01 & $138_{-57}^{+38}$ & $1.67_{-0.08}^{+0.08}$ & 0.14$_{-0.08}^{+0.04}$ & 4437$_{-3994}^{+39936}$ & XMM/INT/BAT & 2 \\
MCG-02-58-022 & 0.047 & $7.1\pm0.6$ & 10 & 1.8 & $>510$ & $1.95_{-0.04}^{+0.03}$ &  $>0.17$ & 1360$_{-1018}^{+4054}$ & XMM/INT/BAT & 2 \\
\hline
 \end{tabular}
\end{center}
\begin{flushleft} 
$F_x$ is the 0.1-200\,keV X-ray flux in
$10^{-10}\,{\rm erg}\,{\rm cm}^{-2}\,{\rm s}^{-1}$.\\
\textbf{References:} 1:\citet{Masetti2009}, 2:\citet{Middleton2007}, 3:\citet{Winter2009}, 4:\citet{Masetti2006}, 5:\citet{Peterson2004}, 6:\citet{Hicks2008}, 7:\citet{Uttley2005}, 8:\citet{Emmanoulopoulos2014}, 9:\citet{Beckmann2009}, 10:\citet{Woo2002} 
\end{flushleft} 
\end{table*}

\begin{table*}
\caption{The targets and properties of the active galactic nuclei analyzed by \citet{Vasudevan2013} included in this work. The reference to the works providing the mass estimates are given in the right-most column.}\label{data_ranjan}
\begin{center}
\begin{tabular}{|c|c|c|c|c|c|c|c|c|c|c|}
\hline Source & z & $\log(M)$ & $r_{\rm co}$ & $F_x$ & $E_{\rm cut}$ & $\Gamma$ & $\Theta$ & $\ell$ & Data & References \\
       &   &  [$M_\odot$] & [$r_G$] &  & [keV] & & & \\
\hline \hline
NGC\,3227 & 0.004 & $6.78_{-0.11}^{+0.08}$ & 10 & 0.28E & $>636$ & $2.08^{+0.05}_{-0.09}$ & $>0.21$  & 3$_{-1}^{+1}$ & XMM/BAT & $^{b}$,1 \\
PG\,1114+445 & 0.144 & $8.4\pm1$ & 10 & 0.080 & $69_{-47}$ & $1.60^{+0.60}_{-0.22}$ & 0.07$_{-0.05}$ & 27$_{-25}^{+245}$ & XMM/BAT & 2 \\
NGC\,4051 & 0.002 & $6.13_{-0.16}^{+0.12}$ & 10 & 0.15 & $>381$ & $2.49^{+0.11}_{-0.12}$ & $>0.12$ & 3$_{-1}^{+1}$ & XMM/BAT & $^{b}$,3 \\
PG\,1202+281 & 0.165 & $8.5\pm1$ & 10 & 0.11 & $556_{-492}$ & $2.08^{+0.17}_{-0.16}$ & $>0.50$ & 46$_{-41}^{+414}$ & XMM/BAT & 2 \\
NGC\,4138 & 0.003 & $6.8_{-0.5}^{+0.5}$ & 10 & 0.46 & $148_{-73}$ & $1.51^{+0.34}_{-0.20}$ & 0.15$_{-0.10}$ & 3$_{-2}^{+6}$ & XMM/BAT & 4\\
NGC\,4151 & 0.003 & $7.56_{-0.05}^{+0.05}$ & 10 & 1.9 & $79^{+4}_{-4}$ & $1.31^{+0.03}_{-0.03}$ & 0.077$_{-0.028}^{+0.003}$ & 2.5$_{-0.3}^{+0.3}$ & XMM/BAT & $5-7$ \\
Mrk\,766 & 0.013& $6.82_{-0.06}^{+0.05}$ & 3.4 & 0.12 & $21^{+7}_{-6}$ & $1.56^{+0.09}_{-0.08}$ & 0.021$_{-0.011}^{+0.007}$ & 13$_{-1}^{+2}$ & XMM/BAT & $8-10$ \\
NGC\,4258 & 0.002 & $7.59_{-0.01}^{+0.01}$ & 10 & 0.59 & $>284$ & $1.82^{+0.06}_{-0.10}$ & $>0.09$ & 0.145$_{-0.003}^{+0.003}$ & XMM/BAT & 11 \\
Mrk\,50 & 0.023 & $7.42_{-0.07}^{+0.06}$ & 10 & 0.52 & $>334$ & $2.18^{+0.02}_{-0.02}$ & $>0.11$ & 46$_{-6}^{+8}$ & XMM/BAT & 12 \\
NGC\,4388 & 0.008 & $7.2_{-0.6}^{+0.6}$ & 10 & 0.14 & $>2096$ & $1.81^{+0.03}_{-0.05}$ & $>0.68$ & 26$_{-20}^{+78}$ & XMM/BAT & 13\\
NGC\,4395 & 0.001 & $5.45_{-0.15}^{+0.13}$ & 51.5 & 0.20 & $45^{+27}_{-10}$ & $1.18^{+0.11}_{-0.12}$ & 0.04$_{-0.02}^{+0.03}$ & 3$_{-1}^{+1}$  & XMM/BAT & 10,14 \\
NGC\,4593 & 0.009 & $6.88_{-0.10}^{+0.08}$ & 10 & 1.7 & $>517$ & $1.89^{+0.08}_{-0.05}$ & $>0.17$ & 76$_{-13}^{+21}$ & XMM/BAT & $15-16$ \\
NGC\,5252 & 0.023 & $9.03_{-0.02}^{+0.40}$ & 10 & 0.50 & $111^{+58}_{-18}$ & $1.38^{+0.09}_{-0.05}$ & 0.11$_{-0.05}^{+0.06}$  & 1.1$_{-0.6}^{+0.1}$ & XMM/BAT & 17 \\
NGC\,5506 & 0.006 & $6.7\pm0.7$ & 10 & 3.2 & $166^{+107}_{-30}$ & $1.85^{+0.02}_{-0.10}$ & 0.16$_{-0.07}^{+0.11}$ & 104$_{-83}^{+417}$ & XMM/BAT & 13\\
NGC\,5548 & 0.017 & $7.59_{-0.21}^{+0.24}$ & 4.5 & 1.8 & $415^{+827}_{-178}$ & $1.73^{+0.02}_{-0.02}$ &  0.41$_{-0.25}^{+0.81}$ & 59$_{-25}^{+37}$ & XMM/BAT & 10,18 \\
Mrk\,1383 & 0.087 & $9.00_{-0.16}^{+0.11}$ & 10 & 0.36 & $>134$ & $2.20^{+0.24}_{-0.20}$ & $>0.04$ & 11$_{-2}^{+5}$ & XMM/BAT & 5 \\
Mrk\,817 & 0.031 & $7.59_{-0.06}^{+0.07}$ & 10 & 0.84 & $>150$ & $2.37^{+0.04}_{-0.09}$ & $>0.05$ & 92$_{-13}^{+17}$ & XMM/BAT & 1,5 \\
Mrk\,841 & 0.036 & $8.5\pm0.7$ & 10 & 0.56 & $>597$ & $2.26^{+0.10}_{-0.09}$ &  $>0.20$ & 10$_{-8}^{+40}$ & XMM/BAT & 19 \\
\hline
 \end{tabular}
\end{center}
\begin{flushleft}
$F_x$ is the 0.1-200\,keV X-ray flux in $10^{-10}\,{\rm erg}\,{\rm
  cm}^{-2}\,{\rm s}^{-1}$.\\
$^{b}$ Reveberation results summarized on Misty Bentz's website (http://www.astro.gsu.edu/AGNmass/).\\
\textbf{References:} 1:\citet{Denney2010}, 2:\citet{Baskin2005}, 3:\citet{Denney2009}, 4:\citet{Winter2009}, 5:\citet{Peterson2004}, 6:\citet{Bentz2006}, 7:\citet{Metzroth2006}, 8:\citet{Bentz2009}, 9:\citet{Bentz2010}, 10:\citet{Emmanoulopoulos2014}, 11:\citet{Uttley2005}, 12:\citet{Barth2011}, 13:\citet{Bian2007}, 14:\citet{Peterson2005}, 15:\citet{Denney2006}, 16:\citet{Barth2013}, 17:\citet{Graham2013a}, 18:\citet{Pancoast2014}, 19:\citet{Middleton2007}
\end{flushleft}
\end{table*}

\begin{table*}
\caption{Microlensing results on the coronae of active galactic nuclei from \textit{Chandra} observations. The references to the individual works are given in the right-most column.}\label{data_sample_lenses}
\begin{center}
\begin{tabular}{|c|c|c|c|c|c|c|c|}
\hline
Source & z & $\log(r_{\rm co})$ & $L$ [0.2-50\,keV] & $E_{\rm cut}$ & $\Theta$ & $\ell$ &  References \\
       &   & [cm] & [$10^{-10}\,{\rm erg}\,{\rm s}^{-1}$] & [keV] & \\
\hline \hline QJ0158-4325 & 1.29 & 14.3 & 5.09E44 & 18 & $>0.02$ & $>69$ & \citet{Chen2012,Morgan2012} \\
HE 0435-1223 & 1.69 & 14.8 & 5.03E44 & 22 & $>0.02$ & $>22$ & \citet{Chen2012,Blackburne2014} \\
SDSS J0924+0219 & 15 & 4.58 & 2.24E44 & 19 & $>0.02$ & $>6$ & \citet{Chen2012,MacLeod2015} \\
HE 1104-1805 & 2.32 & 15.33 & 15.6E44 & 23 & $>0.02$ & $>20$ & \citet{Chen2012,Blackburne2014a} \\
Q2237+0305 & 1.69 & 15.46 & 7.39E44 & 19 & $>0.02$ & $>7$ & \citet{Chen2012,Mosquera2013} \\
\hline
 \end{tabular}
\end{center}
\end{table*}

\end{document}